\documentclass{epl}

\title{
 The effective mass and the $g$-factor of the
strongly-correlated  2-D electron fluid.
Evidence for a coupled-valley state in the Si system.
}

\shorttitle{The effective mass and the $g$-factor}

\author{
M.W.C. Dharma-wardana\thanks{E-mail:\email{chandre@argos.phy.nrc.ca}}}

\institute{Institute of Microstructural Sciences,
National Research Council of Canada, Ottawa,Canada. K1A 0R6\\
}

\begin{document}
\maketitle

\begin{abstract}
The  effective mass $m^*$, and
the  Land\'e $g$-factor  of the
uniform 2-D electron system (2DES) are calculated as a function of the
spin polarization $\zeta$, and the density parameter $r_s$, using a
non-perturbative  analytic approach.
 Our  theory is
in  good accord with the susceptibility data
 for the simple 2DES, and in excellent agreement 
 with the two-valley Si-2DES data of Shashkin et al.
 While $g^*$ is enhanced in GaAs,
$m^*$ is enhanced in Si. The two-valley susceptibility is
 treated within a coupled-mode (coupled-valley)
approach.
The coupled-valley
model is confirmed by comparison with the Quantum Monte Carlo
results for a 4-component 2DES.
\end{abstract}
%
%
%
{\it Introduction.--}
The 2-D electron fluid (2DES) exhibits
a wealth of
intriguing physics, straddling a rich
phase diagram\cite{pd2d,atta}.
The phase diagram  contains 
spin-polarized states at
sufficiently large $r_s$, say $\sim 20-27$. 
Here $r_s=(\pi n)^{-1/2}$ is the electron-disk radius\cite{rscalc,lfc2vpaper}
at the density $n$, in atomic units.
It is  also equal to the value of the coupling
constant $\Gamma$ = (potential energy)/(kinetic energy).
The intermediate
 regime  $r_s$ $\sim 5- 20$ also
 hosts
many ill-understood phenomena including  the metal-insulator
 transition (MIT)\cite{krav}.
Anomalous values
(e.g, see~\cite{zhu}),
of $g^*$ and $m^*$ have been found.
 Some experiments suggest that
an enhancement of
$g^*$ is responsible for the strong enhancement of $m^*g^*$,
while  results\cite{shash} on Si metal-oxide field effect transistors
(MOSFETs) suggest that it is $m^*$, and
not $g^*$ which is enhanced.
In this study we show that, for ideally thin 2-D layers.
$g^*$ is enhanced in GaAs-like systems, while  $m^*$ 
is enhanced in Si-like multi-valley systems.
The existence of a coupled-valley state follows naturally
from the physics of the Si system, and here we present a model
leading to 
excellent quantitative agreement with experiment, and with
Quantum Monte Carlo (QMC) simulations of a 4-component 2DES\cite{cs}.

Fermi liquid-type theories\cite{quinn} are
valid for $r_s<1$. Such perturbative methods have been applied,
invoking impurities\cite{morawetz}, or charge and spin-density
wave effects\cite{galitski}.
On the other hand, 
QMC calculations of $m^*$ 
involve the
{\it excited states} of the 2DES and are less reliable than for
the ground state.
 QMC results up to $r_s=5$ have
been reported~\cite{qmcmstr}.

We showed recently that
the 2-DES, 3-DES, 
and  
dense  hydrogen can be studied using a mapping 
to a classical fluid 
\cite{prl1,prb00,prl2,hyd}. The accuracy of the map
was established by comparison with QMC and other independent
calculations.
 Here
we use this classical map to evaluate $m^*$ and $g^*$ for the
low-density 2-DES.
The method is best understood within a density-functional picture.

{\it The  density-functional perspective.--}
The Hohenberg-Kohn-Mermin theorem asserts that 
the Helmholtz free energy $F$ is a minimum at
 the true density~\cite{hkm}.
If $n(r)$ is
the true density, it obeys the variational equation 
$\delta F[n(r)]/\delta n(r) =0.$   
If the origin of coordinates is on an electron,
then if $n(r)$ is the density as seen
from this electron, it is a {\it pair-density} such that
$n(r)=n\,g(r).$
Here $g(r)$ is the electron-pair distribution function (PDF).
The variational condition gives the  
 Kohn-Sham~(KS) equation as usual. Then $n(r)$ is obtained
via a sum over the KS orbital-densities  $|\psi_i|^2$ weighted by the
 Fermi factors $f_i$.
If the electrons formed a {\it classical system}, the variational equation 
becomes the Boltzmann form for the density:
\begin{equation}
\label{boltzman}
n(r)=ne^{-\beta\{V_{cou}(r)+V_p(r)+V_{c}(r)\}}.
\end{equation}
 $V_{cou}(r)$ is the Coulomb interaction  between the electron at the
origin and the electron located at $\vec{r}$. 
Similarly, $V_p(r)$ is the Poisson potential  
at $\vec{r}$, and   
$V_{c}(r)$ is a correlation potential. For a 
classical system  the $V_{xc}(r)$ of standard KS theory is replaced 
by just a correlation potential $V_c(r)$.
In effect,Eq.~\ref{boltzman} evaluates
 the $g(r)$ of the classical
fluid. However, the $g(r)$ of a classical fluid is accurately
given by the hyper-netted chain (HNC)
inclusive of a bridge function\cite{hncref}. Thus, the
extended HNC equation is a classical KS equation where
$V_c(r)$ is the sum of HNC+bridge
diagrams. The construction of the Bridge diagrams for the
2DES is given in refs.\cite{prl2,totsuji}.

The classical map has no exchange, and fails
as $T\to 0$. We rectify these  lacunae as follows.
In a system {\it without} Coulomb
interactions,  $g(r)$ should reduce to $g^0(r)$ which is known
analytically (at $T=0$) or numerically.  The first
step of the mapping is to introduce a potential $\phi^0_{ij}(r)$ (where $i$,
 $j$ are spion labels)
such that $\phi^0_{ij}(r)$
generates $g^0_{ij}(r)$ when used in the HNC equation for ideal
electrons~\cite{lado1}.
This leads to an exact
treatment of exchange.

Electrons at $T=0$ have kinetic energy.
Hence the
classical map of the quantum fluid at $T=0$ would be at some 
``quantum temperature'' $T_q$. This is determined by requiring the
correlation energy $\epsilon_c$ of the classical fluid at $T_q$  be
 equal to the
$\epsilon_c$ of the quantum fluid at $T=0$. This may be regarded as
a "calibration" of the classical fluid to recover the
 quantum exchange-correlation
energy in the $r_s$ range of interest.
Here we use the $\epsilon_c(r_s)$ given by QMC
 (Tanatar-Ceperley results for the fully spin-polarized 2DES for $r_s$ up tp 30
 were used in~\cite{prl2}). 
Once $T_q$, which maps the $T=0$ quantum fluid to a classical fluid
 is known, finite-$T$
fluids are calculated from classical fluids at the temperature
 $T_{cf}=(T^2+T_q^2)^{1/2}$, as justified in ref.\cite{prb00}.
We have shown\cite{prl1,prl2} that the classical PDFs are in
very  close agreement with
 the quantum fluid PDFs obtained via QMC.
The success of the method(refs.\cite{prl1,prb00,prl2,hyd})
 for 2-D and 3-D electrons, hydrogen fluids, and for
4-component 2-D electron fluids as judged by comparison with QMC data
establishes it to be a well controlled, highly reliable method.
The PDFs are easily 
 used in a coupling-constant integration
for the exchange-correlation free energies $F_{xc}$.
Our finite-$T$ method accurately recovers the low-$T$
logarithmic terms in $F_c$ which cancel with corresponding terms in $F_x$.
This method, based on a {\it classical} mapping of the quantum calculation to
an HNC calculation is called CHNC
~\cite{pd2d,prl1,prb00,prl2}.

{\it Evaluation of  $m^*$ and  $g^*$.--}
	  The evaluation of the susceptibility enhancement
$m^*g^*$ uses the $T=0$ results
 for the exchange-correlation energy
 $\epsilon_{xc}(r_s,\zeta)$. This is expressed in terms of 
 $\epsilon_{xc}(r_s,0)$ and $\epsilon_{xc}(r_s,1)$, and a polarization
 factor $P(r_s,\zeta)$ given in Eq.~(6) of Ref.~\cite{prl2}.
Using Hartree units, the ratio of the static spin susceptibility to
 the ideal (Pauli) spin susceptibility is:
\begin{equation}
\label{chi}
\chi_P/\chi_s= (m^*g^*)^{-1} = 1+
 r_s^2\,\partial^2 \epsilon_{xc}/\partial \zeta^2 .
\end{equation}
The effective mass $m^*$ at temperature $T$ is the ratio of the
specific heats, $C_v(T)/C^0_v(T)$ of the interacting and non-interacting 2DES.
\begin{equation}
\label{mstareq}
m^*(T)=\, C_V(r_s,\zeta,T)/C_0(r_s,\zeta,T).
\end{equation}
The specific heats are obtained as the second-T derivatives of
the interacting and ideal Helmholtz $F(r_s,T)$. Here $T$ is the physical
temperature and not $T_{cf}$. The latter is used only in the classical map
to obtain the PDFs.
$F_x(r_s,T)$ has a logarithmic term of the form $T^2\log(T)$ which is
cancelled by a similar term in $F_c(r_s,T)$. That is,
\begin{eqnarray}
\label{logform}
F_x&=&A_x+B_x t^2\log(t)-C_xt^2,\;\, t=T/E_F\\
F_c&=&A_c+B_c t^2\log(t)-C_xt^2,\;\, B_x=-B_c.
\end{eqnarray}
This cancellation holds to 85-95\% in our numerical
CHNC results, for the range $r_s=5-30$, $0<t<0.25$.
 Thus, at $r_s=15$ and 25,
 $(B_x,B_c)$ are (-0.0258, 0.0228), and (-0.0155, 0.0142).
If Hubbard-type finite-$T$
 RPA  were used in the self-energy,
the cancellation is quite poor, even at low-$r_s$. 
These logarithmic terms and the $m^*$ have
also been studied by Geldart et al., using CHNC
\cite{wally}.
%
{\it Multi-valley systems-}
 Shashkin et al.\cite{shash}, also\cite{pudalov},
 have studied {\it clean} low-density 2-valley 2DES in Si-MOSFETs. 
The two valleys are assumed degenerate\cite{valpud}.
It is found\cite{shash} that
the  $m^*$ is strongly enhanced, while
$g^*$ shows little change. The enhanced $m^*$ is 
{\it independent} of $\zeta$. These results, 
''contrary to normal expectations'',
are reproduced by 
 our coupled-mode theory of two valleys.
\begin{figure}
\includegraphics*[width=6.5 cm, height=8 cm]{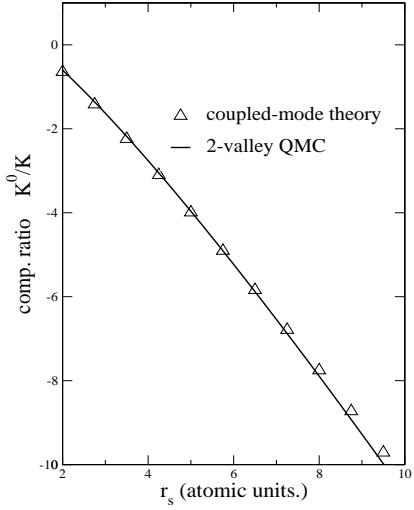}
\caption{
The compressibility ratio $K^0/K$ calculated from the
4-component (2-valley) QMC results of Conti et al., and from the
2-component (1-valley) data using the $k\to 0$ limit of
$\Pi_0/\Pi_{cm}$ where $\Pi_{cm}$ is  
the coupled-mode  polarization function. 
}
\label{k0kfig}
\end{figure}

%

Two equivalent valleys and  two spins imply 
10 different PDFs,
 $g^{uv}_{ij}$, where $u,v$ 
are valley indices.
Such a calculation for each $r_s, \zeta$, $T$  and 
many values of the coupling constant is 
laborious. A simpler
procedure using just
three PDFs is possible. Even if $\zeta\neq 0$, each valley has a density
$n/2$. Thus the 2-valley system may be made up from the known
properties of the one-valley (two-spin) 2DESs coupled together
by their Coulomb interaction. The individual 1-valley correlation
free energies
$F_c^u$, $F_c^v$ are known from QMC and CHNC results.
The inter-valley term for a system with a total density $n$,
 and valley densities $n/2$ is {\it not} known.
Here we present a {\it simple approximation} validated
by calculating the 2-valley compressibility in the same way and comparing 
 with the QMC data of Conti et al\cite{cs}.
There is no exchange interaction between up-spin and
down-spin electrons in the one-valley system,
and the spin densities are $n/2$ at $\zeta=0$.
Hence, since $F_c^u(n/2)$, $F_c^v(n/2)$, and $F_c(n,\zeta=0,[g_{12}])$
for the one-valley system are known, we build up the
2-valley system within the assumption that 
$F_c(n,\zeta=0)$ can be used for the inter-valley contribution
to the $F_c$ of the 4-component (i.e., 2-valley) system.
In a full 4-component CHNC calculation,
the inter-valley interaction is switched
on via a coupling constant integration.
This effect can be recovered
within linear response
by developing the coupled-mode 2-valley response functions.
An analogous coupled-mode problem arises in electron-hole
systems (see Vashishta et al\cite{vbs}).

The total (spin or charge) density-fluctuation spectrum
 of the electrons in a
given (single) valley $v$ is described by the
 response functions $\chi_v=\chi_v^0/D_v$,
where $\chi_v^0$ is the 2-D Lindhard function  
 weighted 
appropriately with the square of the Bohr magneton $\mu_B$ or unity, and
$D_v$ is a corresponding denominator for each case. The Pauli
susceptibility $\chi_P$ is the long-wavelength limit $\mu_B^2\chi_v^0(k=0)$.
Let us consider a denominator of a response function (which may be
the charge response $\chi$, the proper polarization function $\Pi$,
 or the spin susceptibility
$\chi_s$, depending on how the local-field factor $G_v$
is specified). The  
 denominator $D_v=1-v_{cou}(1-G_v)\chi_v^0$ and defines  $G_v$,
 the local-field
factor (LFF, see \cite{deflfc}). We are only concerned with the 
static $k\to 0$ limit. Then $G_v$ for $\Pi$ are related to
$K^0/K$, while the $G_v$
 for $\chi_s$ is given by  $\chi_P/\chi_s$, as in Eq.~\ref{chi},
 and depends on the correlation free energy $F_c$ of the one-valley 2DES.
 When two such 2DESs,
described by $\chi_v$ and $\chi_u$
interact {\it via} the inter-valley term, 
{\it coupled modes} are formed. These
 modes are described by the zeros of a {\it  new denominator}
of the response function of the {\it total} 2-valley  system. 
This coupled-mode form is\cite{vbs}:
\begin{eqnarray}
\chi_{cm}&=&[\chi_u^0+\chi_v^0+v_{cou}^2\chi_u^0\chi_v^0(\Sigma G_{uv})]/D_{cm}\\
\Sigma G_{uv}&=&G_u+G_v
-G_{uv}-G_{vu}\\
D_{cm}&=&D_uD_v-v_{cou}^2\chi_u^0\chi_v^0(1-G_{uv})(1-G_{vu})
\end{eqnarray}
Here $G_{uv}$ is an LFF arising from the inter-valley term 
$F_{uv}$ already discussed, and modeled by $F_{12}(n,\zeta=0)$ at $k=0$.
 Hence we express the susceptibility enhancement $\chi_s/\chi_P$ as
$\chi_{cm}/\chi_P$, and this is evaluated from the $G_u,G_v$ and $G_{uv}$.
 Equation~\ref{chi} determines $G_u$ = $G_v$, where the correlation part
involves the  
second derivative ($r^2_{sv} d^2F_c^v/d\zeta^2$). Similarly 
the cross term $G_{uv}$ involves $r^2_{s} d^2F^c_{12}(n,\zeta=0)/d\zeta^2$.
The 4-component QMC results of Ref.~\cite{cs} for $F_c(r_s,\zeta=0,T=0)$
enable us to calculate the compressibility ratio $K^0/K$ of the 2-valley
system directly. The coupled-mode theory, applied to the
 proper polarization function $\Pi$
gives another evaluation $K^0/K$.
The agreement between the two methods
is shown in Fig.~\ref{k0kfig}. A similar comparison for $\chi_P/\chi_s$
is not possible as the QMC results are available  only at $\zeta=0$.
 However, the
agreement between the two estimates of $K^0/K$
validates our coupled-mode evaluation of 2-valley properties
from the 1-valley  energies.
Thus the 2-valley results are constructed from the 1-valley CHNC energies
(which argee closely with QMC data) which include the usual
bridge contributions\cite{prl2}.

\begin{figure}
\includegraphics*[width=8 cm, height=10.0 cm]{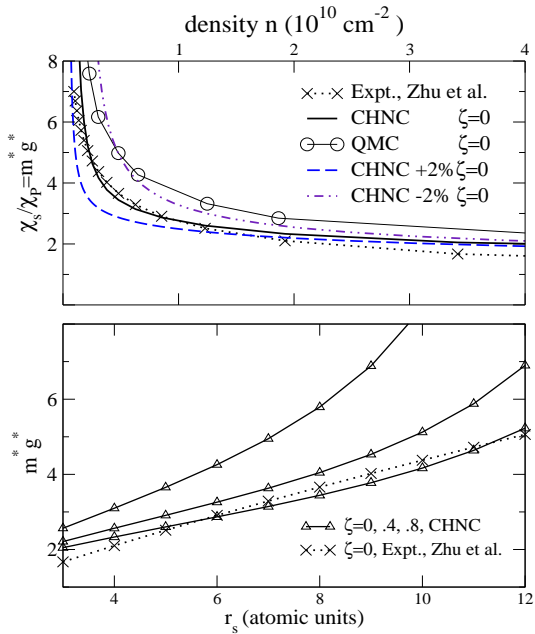}
\caption
{ The spin-susceptibility enhancement $\chi_s/\chi_P$ = $m^*g^*$ in the 2DES.
 Top panel: comparison of 
experiment~\cite{zhu}, QMC~\cite{atta}. and CHNC. Curves marked $\pm$2\%
are CHNC predictions if the exchange-correlation contribution
$r_s^2\,\partial^2 \epsilon_{xc}/\partial \zeta^2$ is modified by $\pm2$\%.
Bottom panel:
 CHNC results for $m^*g^*$
for 3 spin-polarization $\zeta$, and the experimental $\zeta$=0 data,\
 plotted against $r_s$.
}
\label{figzhu}
\end{figure}

{\it Results--}
 In Fig.~\ref{figzhu} we show $\chi_s/\chi_P=m^*g^*$ for a
single-valley system,
 as a function of the density $n$, and as
a  function of $r_s$ (see \cite{rscalc})
at $T=0$ for $\zeta=0$.
 Our results, the experimental data of Zhu et al.~\cite{zhu},
 and QMC 
data, extracted from Fig. 2  of ref.~\cite{atta}
 are displayed.
The high density regime\cite{colm} is in agreement with
standard theories and is not displayed.

 For the Zhu et al. data we use their fitted form
$m^*g^* = (2.73+3.9\,n\,\zeta)\,n^{-0.4}$
 where the density $n$ is in units of 10$^{10}$ cm$^{-2}$.
The strong agreement between CHNC and the Zhu data is perhaps
fortuitous since the results are quite
sensitivity to the $d^2/d\zeta^2$ calculation
to the energy
differences $\Delta E= E_c(\zeta=1)-E_c(\zeta=0)$
and the form of the polarization factor $P(r_s,\zeta)$.
The CHNC is calibrated to the Tenatar-Ceperley QMC
which differs somewhat from the Attaccalite data. We have also
plotted two CHNC curves where (see Eq.\ref{chi}) the term
 $r_s^2\,\partial^2 \epsilon_{xc}/\partial \zeta^2$ has been modified by
$\pm$ 2\%. Clearly, errors in converting to $r_s$, modification
of exchange-correlation gradients by well-width effects
and the presence of impurities
etc., can produce such a change.
 The bottom panel (Fig.~\ref{figzhu}) 
 shows the comparison against
$r_s$. 

Zhu et al. report a $\zeta$ dependence, 
but now they
consider that the finiteness of the 2-D layer and orbital effects
cannot be
ignored
in analysing field-dependent data\cite{zhunara}.
As pointed out via the $\pm2\%$ plots in Fig.\ref{figzhu},
 the sensitivity of $\chi_P/\chi_s$ to small
errors in the xc-energy gradient is also important.
(Discussion of these and
 other data for $m^*$ and $g^*$,
 of Zhu's thesis\cite{zhu} will await their publication).

In our results,
$\chi_s/\chi_P$ is less sensitive to $\zeta$ at high density,
 and very sensitive to $\zeta$ at low density, approaching the 
 {\it para}$\to${\it ferro} transition. In fact, the second derivative
 in Eq.\ref{chi} diverges at $\zeta=1$.
\begin{figure}
\includegraphics*[width=8.5 cm, height=9.5 cm]{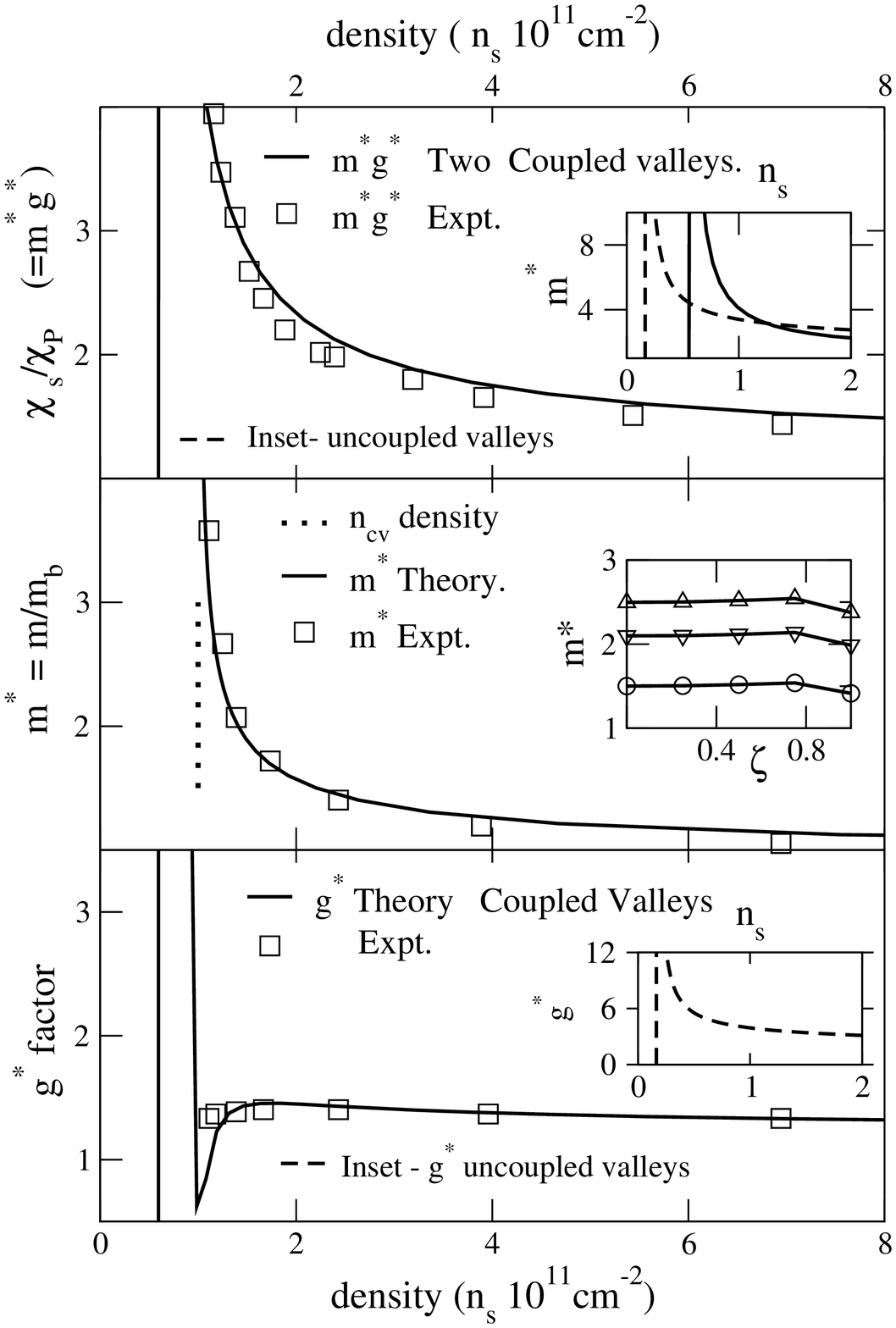}
\caption{
Comparison of experiment\cite{shash} and theory for the 2-valley 2DES in
Si-MOSFETs.
The top panel shows $m^*g^*$, while the
 inset shows the shift of theoretical $m^*g^*$ curve to higher densities
due to mode coupling. The middle panel shows $m^*$ which rises steeply at the
onset of the spin-singlet coupled-valley state at $n_{cv}$=1x10$^{11}$/cm$^2$. The
inset shows the insensitivity of $m^*$ to the spin-polarization for three densities.
The bottom panel compares the experimental $g^*$ with
theory.
}
\label{simg}
\end{figure}

A very different
 experimental
picture is found in Si-2DESs~\cite{shash}.
The CHNC results for the coupled  2-valley 2DES are shown in
Fig.~\ref{simg}. The top panel compares the $m^*g^*=\chi_{cm}/\chi_P$
obtained from experiment and  the coupled-mode analysis (the LFFS used
are for the spin-spin response).
 The inset shows 
the shift of the simple uncoupled-valley  curve to higher densities
when the valley coupling is introduced.
The conversion between density and $r_s$
is discussed in ref.~\cite{rscalc}.

 The middle panel (fig.~\ref{simg}) shows the $m^*$ calculated from the
finite-$T$ analysis, with the sharp rise occurring 
at $r_s\sim 5.4$,  i.e., density
$n_{cv}$ = 1x10$^{11}$/cm$^2$.
The inset shows the lack of $\zeta$ dependence in $m^*$
for three densities.
This is because the physics is dominated by singlet interactions,
 as in the
 ambi-spin phase reported earlier\cite{pd2d}.
The lower panel of Fig.~\ref{simg} shows the flat $g^*$ of the coupled-valley
fluid, while the inset shows the usual  increase of $g^*$ 
 in the
uncoupled system as the density is reduced.

{\it Conclusion--}
	We have presented results
for the effective mass  $m^*$, and the 
Land\'{e} $g^*$ factor
 of  2-D electron
fluids, using an analytic method.  Our results suggest that
exchange effects dominate as $r_s$  increases in 1-valley 2D  system,
enhancing $g^*$ when the one-valley spin-response diverges.
Correlation effects outweigh exchange
in 2-valley systems where  $m^*$ is strongly
enhanced and only weakly dependent on $\zeta$.
The tendency to form singlets already noted in the
single 2DES\cite{lfc} becomes stronger in the 2-valley 2DES
where a coupled-valley state is formed.
 Our theoretical results depend only on the 
 $\epsilon _{xc}(r_s,\zeta,T)$ used
in calculating $\zeta$ and $T$ derivatives,
and invoke no fit parameters specific to this problem.
However, the inter-valley energy $F_{uv}$ was approximated 
via the $F_{12}(\zeta=0)$ of the one-valley inter-spin
energy. As already noted, $\chi/\chi_0$ is 
quite  sensitive
to the evaluation of $d^2/d\zeta^2$.
However, the agreement
of the present model with experiment may prove useful
in understanding the experimental results.

\acknowledgments We thank Peter Coleridge, Wally Geldart,
 Fran\c{c}ois Perrot, Sasha Shashkin,
 Horst Stormer and Jun Zhu for their comments and correspondence.
\newpage

\begin{thebibliography}{0}
\bibitem{deflfc}
\Publ{Unlike in ref.\cite{quinn} we use the same form, $1-v_{cou}(1-G_v)$,
for both
symmetric and antisymmetric LFFs, using the same write up
for both. 
Thus $G_{-}$ of ref.\cite{quinn} is our $G_v-1$
for the spin-$\chi$,
while their $G_+$ is our $G_v$ for the density-$\chi$.}
\end{thebibliography}
\end{document}